\title{On the self-force in electrodynamics and implications for gravity}

\author{Volker Perlick\footnote{Email: perlick@zarm.uni-bremen.de} \\
        ZARM, University of Bremen\\
        Am Fallturm, 28359 Bremen, Germany}

\date{}

\documentclass[11pt,english]{article}
\usepackage{amsmath,amssymb}
\usepackage{amsthm}        
\usepackage{cancel}
\usepackage{graphics,epsfig,psfrag}

\begin{document}
\maketitle

\begin{abstract}
We consider the motion of charged point particles on Minkowski
spacetime. The questions of whether the self-force is finite 
and whether mass renormalisation is necessary are discussed
within three theories: In the standard Maxwell vacuum theory,
in the non-linear Born-Infeld theory and in the higher-order  
Bopp-Podolsky theory. In a final section we comment
on possible implications for the theory of the self-force
in gravity.
\end{abstract}

\section{Introduction}
\label{sec:intro}
The problem of the electromagnetic self-force has a long
history. It began in the late 19th century when Lorentz,
Abraham and others tried to formulate a classical theory
of the electron. The idea was to model the electron as 
an extended, at least approximately spherical, charged
body and to determine the equations of motion for the
electron. Based on earlier results by Lorentz, Abraham 
succeeded in writing the equation of motion in terms of
a power series with respect to the radius of the electron. 
If the radius tended to zero, i.e., for a point charge, an 
infinity occurred. The reason for this infinity is in the 
fact that, in the point-particle limit, the electric field 
strength diverges so strongly at the position of the charge
that the field energy in an arbitrarily small sphere 
becomes infinitely large. To get rid of
this infinity, it was necessary to ``renormalise the mass''
of the particle by assuming that it carries a negative
infinite ``bare mass''. After this mass renormalisation,
one got a differential equation of third order for the
motion of the particle which is known as the Abraham-Lorentz
equation. It is a non-relativistic equation in the sense
that, on the basis of special relativity, it can hold only
if the particle's speed is small in comparison to the
speed of light.

A fully relativistic treatment of the problem had to wait
until  Dirac's work \cite{Dirac:1938} of 1938. The resulting 
equation of motion is known as the Lorentz-Dirac equation
or as the Abraham-Lorentz-Dirac equation. Clearly, everyone
would call it the Dirac equation except for the fact that this
name was already occupied by another, even more famous 
equation. Neither Lorentz nor Abraham has ever seen the
(Abraham-)Lorentz-Dirac equation, because both had passed away 
in the 1920s. In particular in the case of Abraham it is rather 
clear that he would not have liked this equation because he 
was an ardent opponent of relativity. Therefore, it seems 
appropriate to omit his name and call it the Lorentz-Dirac 
equation. For the derivation of the Lorentz-Dirac equation,
again mass renormalisation was necessary and one arrived at a 
third-order equation of motion. The latter fact means that, in 
contrast to other equations of motion, not only the position and 
the velocity but also the acceleration of the particle has to be 
prescribed at an initial instant for fixing a unique solution. 
Moreover, the Lorentz-Dirac equation is notorious for showing
unphysical behaviour such as run-away solutions and pre-acceleration.
For a detailed discussion of the Lorentz-Dirac equation, including
historical issues, we refer to Rohrlich~\cite{Rohrlich:2007}
and to  Spohn~\cite{Spohn:2007}.

The trouble with the Lorentz-Dirac equation clearly has its
origin in the fact that the electric field strength of a point
charge becomes infinite at the position of the charge, and that
this singularity is so strong that the field energy in an arbitrarily
small ball around the charge is infinite. A possible way out is to
modify the underlying vacuum Maxwell theory in such a way that this 
field energy becomes finite. Two such modified
vacuum Maxwell theories have been suggested in the course of
history, the non-linear Born-Infeld theory \cite{BornInfeld:1934}
and the linear but higher-order Bopp-Podolsky theory 
\cite{Bopp:1940,Podolsky:1942}. It is the main purpose of this 
article to discuss to what extent these theories have succeeded
in providing a theory of classical charged point particles with a
finite self-force and a finite  field energy.

Some people are of the opinion that there is no need for a consistent
theory of classical charged point particles. They say that either one
should deal with extended classical charge distributions or with
quantum particles. However, this is not convincing. E.g. in accelerator 
physics it is common to describe beams in terms of classical point 
particles; neither a description in terms of extended charge distributions 
nor in terms of quantum matter seems to be appropriate or even feasible.
Therefore, a consistent and conceptually well-founded theory of classical
charged point particles is actually needed.

The problem of the electromagnetic self-force of a charged particle
has a counterpart in the gravitational self-force of a massive
particle. In comparison with the electromagnetic self-force, the 
gravitational self-force is plagued with additional conceptual
issues. The latter are related to the facts that Einstein's field
equation does not admit solutions for sources concentrated on a 
worldline, see Geroch and Traschen~\cite{GerochTraschen:1987}, and that
an extended massive particle becomes a black hole if it is compressed
beyond its Schwarzschild radius. However, by considering the
self-interacting massive particle as a perturbation of a fixed
background spacetime one arrives at a formalism which is similar
to the electromagnetic case, see the comprehensive review by
Poisson, Pound and Vega~\cite{PoissonPoundVega:2011}. At this 
level of approximation it is reasonable to ask if modifications
of the vacuum Maxwell theory can be mimicked by modifying 
Einstein's theory in such a way that the (approximated) 
gravitational self-force becomes finite. We will come back to
this question at the end of this article, after a detailed
discussion of the electromagnetic case.

\section{Maxwell's equations and the constitutive law for vacuum}\label{sec:Max}

Maxwell's equations are universal and they do not involve a
metric or a connection. They read
\begin{equation}\label{eq:Max}  
dF = 0 \; , \qquad dH=j \; ,
\end{equation}
where $F$ is an untwisted two-form, $H$ is a twisted two-form 
and $j$ is a twisted three-form. (A differential form is twisted
if its sign depends on the choice of an orientation. The difference 
between twisted and untwisted differential forms becomes irrelevant 
if the underlying manifold is oriented.) $F$ gives the electromagnetic 
field strength, $H$ gives the electromagnetic excitation and $j$ 
gives the electromagnetic current. Our notation follows
Hehl and Obukhov~\cite{HehlObukhov:2003}.

The equations (\ref{eq:Max}) are referred to as the \emph{premetric}
form of Maxwell's equations. These equations immediately imply
that on simply conected domains $F$ can be represented in terms
of a potential,
\begin{equation}\label{eq:FA}
F=dA \, ,
\end{equation}
and that charge conservation is guaranteed,
\begin{equation}\label{eq:cont}
d \, j=0 \, .
\end{equation}

If $j$ is given, Maxwell's equations must be supplemented
with a constitutive law relating $F$ and $H$ to specify the
dynamics of the electromagnetic field. There is a particular
constitutive law for vacuum, and there is a particular 
constitutive law for each type of medium. In any case, the
constitutive law will involve some background geometry. In
the following we consider vacuum electrodynamics
on Minkowski spacetime. Then the constitutive law should
involve the Minkowski metric tensor and no other background
fields.

On Minkowski spacetime, we may choose an orthonormal coframe,
i.e., four linearly independent covector fields $\theta ^0, 
\theta ^1, \theta ^2, \theta ^3$ such that the Minkowski
metric is represented as 
\begin{equation}\label{eq:gtheta}
g = \eta _{ab} \, \theta ^a \otimes \theta ^b
\end{equation}
where $( \eta _{ab} ) = \mathrm{diag}(-1,1,1,1)$. Here and
in the following we use the summation convention for latin
indices that take values 0,1,2,3 and for greek indices that
take values 1,2,3. Latin indices will be lowered 
and raised with $\eta _{ab}$ and its inverse $\eta ^{ab}$, 
respectively, while greek indices will be lowered and raised with
the Kronecker symbol $\delta _{\mu \nu}$ and its inverse 
$\delta ^{\mu \nu}$, respectively. 

With respect to the chosen orthonormal coframe, the electromagnetic 
field strength can be decomposed into electric and magnetic parts, 
\begin{equation}\label{eq:EB}
F \, = \,  E_{\mu} \theta ^{\mu} \wedge \theta ^0
\, + \, \dfrac{1}{2} \, B^{\rho} 
\varepsilon _{\rho \mu \nu} \theta ^{\mu}
\wedge \theta ^{\nu} \, .
\end{equation}
Here the wedge denotes the antisymmetrised tensor product and
$\varepsilon _{\rho \mu \nu}$ is the Levi-Civita symbol,
defined by the properties that it is totally antisymmetric
and satisfies $\varepsilon _{123}=1$. The electromagnetic
excitation can be decomposed in a similar fashion,
\begin{equation}\label{eq:DH}
H \, = \, - \, \mathcal{H} _{\mu} \theta ^{\mu} \wedge \theta ^0
\, + \, \dfrac{1}{2} \, D^{\rho} \varepsilon _{\rho \mu \nu} \theta ^{\mu}
\wedge \theta ^{\nu} \, .
\end{equation}
If we apply the Hodge star operator of the Minkowski metric
to $F$ and $H$, we find
\begin{equation}\label{eq:HodgeF}
^*{\!}F \, = \, - \, B_{\mu} \theta ^{\mu} \wedge \theta ^0
\, + \, \dfrac{1}{2} \, E^{\rho} 
\varepsilon _{\rho \mu \nu} \theta ^{\mu}
\wedge \theta ^{\nu} \, ,
\end{equation}
\begin{equation}\label{eq:HodgeH}
^*{\!}H \, = \, - \, D_{\mu} \theta ^{\mu} \wedge \theta ^0
\, - \, \dfrac{1}{2} \, \mathcal{H} ^{\rho} 
\varepsilon _{\rho \mu \nu} \theta ^{\mu}
\wedge \theta ^{\nu} \, .
\end{equation}

The field energy density measured by an observer whose 4-velocity
$V$ satisfies $\theta ^{\mu} (V) =0$ for $\mu = 1,2,3$ is given by
\begin{equation}\label{eq:energy}
\varepsilon \, = \, \dfrac{1}{2} \, 
\big( E_{\mu}D^{\mu}+\mathcal{H}_{\mu}B^{\mu} \big) \, .
\end{equation}

With the help of the Hodge star operator we can form out of
$F$ the untwisted scalar invariant
\begin{equation}\label{eq:I1}
^*{\!}(F \wedge {}^*{\!}F) = B_{\mu}B^{\mu}-E_{\mu}E^{\mu}
\end{equation}
and the twisted scalar invariant
\begin{equation}\label{eq:I2}
^*{\!}(F \wedge F) \, = \, - \, 2\, E_{\mu} B^{\mu} \, .
\end{equation}

All these equations are valid with respect to any orthonormal coframe. 
In particular, we may choose a holonomic coframe, i.e., we may choose 
inertial coordinates on Minkowski spacetimes,
\begin{equation}\label{eq:g}
g = 
\eta _{ab} \, dx^a \otimes dx^b
\end{equation}
and then write $\theta ^a =dx ^a$. In the following we will see
that it is sometimes convenient to work with an anholonomic 
orthonormal coframe on Minkowski spacetime.  

We will now discuss the vacuum constitutive 
law in three different theories.

\subsection{Standard Maxwell vacuum theory}
\label{subsec:Maxcon}

In the standard Maxwell theory, the constitutive law of
vacuum reads
\begin{equation}\label{eq:Maxcon}
H \, = \, {}^*{\!}F \, . 
\end{equation}
By comparison of (\ref{eq:DH}) and (\ref{eq:HodgeF}) we
see that this implies
\begin{equation}\label{eq:Maxcon3}
D^{\rho}=E^{\rho} \, , \qquad \mathcal{H} ^{\mu} = B^{\mu} \, .
\end{equation}
Here and in the following, we use units making the permittivity
of vacuum, $\varepsilon _0$, the permeability of vacuum, $\mu _0$,
and thus the vacuum speed of light, $c=(\varepsilon _0 \mu _0 )^{-1/2}$, 
equal to one.

\subsection{Born-Infeld theory}
\label{subsec:BIcon}
In 1934, Born and  Infeld~\cite{BornInfeld:1934} suggested
a non-linear modification of the vacuum constitutive law,
\begin{equation}\label{eq:BIcon}
H = \dfrac{\, {}^*{\!}F \, - \, \dfrac{^*{\!}(F \wedge F)}{2
b^2} \, F \,}{
\sqrt{\, 1 \, + \, \dfrac{^*{\!}(F \wedge {}^*{\!}F)}{ 
b^2} 
\, - \, 
\dfrac{\big( ^*{\!}(F \wedge F) \big)^2}{4 
b^4} \, } 
}
\end{equation}
where $b$ is a new hypothetical constant of nature with the
dimension of a (magnetic or electric) field strength. The 
idea behind this modified constitutive law is to find a 
theory where the field energy of a point charge
remains bounded. We will discuss in the following sections
to what extent this goal was achieved.

Maxwell's equations with the Born-Infeld constitutive law
(\ref{eq:BIcon}) can be derived from a Lagrangian that 
depends only on the invariants (\ref{eq:I1}) and (\ref{eq:I2}).
This demonstrates that the theory is not only gauge invariant
but also Lorentz invariant. However, we will not need the 
Lagrangian formulation in the following.  
 
As the constitutive law (\ref{eq:BIcon}) does not involve any derivatives,
in the Born-Infeld theory the vacuum Maxwell equations are 
still of first order with respect to the field strength (i.e., 
of second order with respect to the potential), just as in the 
standard Maxwell theory. However, they are now non-linear.

Obviously, the Born-Infeld constitutive law (\ref{eq:BIcon})
approaches the standard vacuum constitutive law (\ref{eq:Maxcon})
in the limit $b \to \infty$. This implies that the Born-Infeld theory
is indistinguishable from the standard Maxwell vacuum theory if $b$ 
is sufficiently big. In this sense, any experiment that confirms
the standard Maxwell vacuum theory is in agreement with Born-Infeld
theory as well, and it gives a lower bound for $b$. For the purpose 
of this article, the specific value of $b$ is irrelevant as long
as it is finite. 

Decomposing the constitutive law (\ref{eq:BIcon}) into
electric and magnetic parts results in
\begin{equation}\label{eq:BIDE}
D^{\rho} = 
\dfrac{
\, E^{\rho} \, + \, \dfrac{E_{\tau}B^{\tau}}{b^2} B^{\rho}
}{
\sqrt{\, 1 + \dfrac{1}{b^2} \big( 
B_{\sigma}B^{\sigma} - E_{\sigma}E^{\sigma} \big)
- \dfrac{\big(E_{\nu}B^{\nu}\big)^2}{b^4} \, } 
} \, ,
\end{equation}
\begin{equation}\label{eq:BIHB}
\mathcal{H} ^{\mu} = 
\dfrac{
\, B^{\mu} \, - \, \dfrac{E_{\tau}B^{\tau}}{b^2} E^{\mu}
}{
\sqrt{\, 1 + \dfrac{1}{b^2} \big( 
B_{\sigma}B^{\sigma} - E_{\sigma}E^{\sigma} \big)
- \dfrac{\big(E_{\nu}B^{\nu}\big)^2}{b^4} \, } 
} \, .
\end{equation}
   
\subsection{Bopp-Podolsky theory}
\label{subsec:BPcon}
Another modification of the vacuum constitutive law, again
motivated by the wish of having the field energy of a point 
charge finite, was brought forward in 1940 by 
Bopp~\cite{Bopp:1940}. The same theory was independently
re-invented two years later by Podolsky~\cite{Podolsky:1942}.
The Bopp-Podolsky theory is equivalent to another theory
that was suggested in 1941 by Land{\'e} and 
Thomas~\cite{LandeThomas:1941}. 

The Bopp-Podolsky vacuum constitutive law reads
\begin{equation}\label{eq:BPcon}
H = {}^*{\!}F - \ell ^2 \square ^*{\!}F
\end{equation}
where  
\begin{equation}\label{eq:box}
\square = {}^*{\!} d ^* {\!} d + 
d ^* {\!} d ^* {\!}
\end{equation}
is the wave operator on Minkowski spacetime and $\ell$
is a new hypothetical constant of nature with the dimension
of a length. In contrast to the Born-Infeld 
constitutive law, the Bopp-Podolsky constitutive law is 
linear. However, it involves second derivatives of the field 
strength, so Maxwell's equations give a system
of fourth-order differential equations for the potential $A$.
In the Land{\'e}-Thomas version of the theory one splits the
potential into two parts each of which satisfies a second-order 
differential equation, see Section~\ref{subsec:BPLW} below.
Just as the Born-Infeld theory, the Bopp-Podolsky can be
derived from a gauge-invariant and Lorentz-invariant Lagrangian
(see Bopp~\cite{Bopp:1940} or Podolsky~\cite{Podolsky:1942})
but we will not use the Lagrangian formulation in the following.

For $\ell \to 0$, the Bopp-Podolsky constitutive law
(\ref{eq:BPcon}) approaches the standard vacuum law
(\ref{eq:Maxcon}). So any experiment that is 
in agreement with the standard Maxwell theory is
in agreement with the Bopp-Podolsky theory as long as
$\ell$ is sufficiently small. However, dealing with the
limit $\ell \to 0$ requires some care because it is a 
\emph{singular} limit of Maxwell's equations 
in the sense that it kills the highest-derivative term.

\section{Field of a static point charge}
\label{sec:static}
It is our goal to discuss the field of a 
point charge in arbitrary motion on Minkowski spacetime 
(subluminal, of course) according to the standard Maxwell 
vacuum theory, the Born-Infeld theory and the Bopp-Podolsky 
theory. As a preparation for that, it is useful to consider
first the simple case of a point charge that is at rest in
the spatial origin of an appropriately chosen inertial 
coordinate system. 
(Obviously, in any other inertial system the charge is then 
in uniform and rectilinear motion.) In this inertial system, 
the field produced by the charge must be spherically symmetric 
and static because there are no background structures that 
could introduce a deviation from these symmetries. Writing
$\vec{r}=(x^1,x^2,x^3)$ for the coordinates and $\vec{E} = 
(E^1,E^2,E^3)$ etc. for the fields in the chosen inertial 
system, Maxwell's equations (\ref{eq:Max})
reduce to 
\begin{equation}\label{eq:Maxstat}
\nabla \times \vec{E} \, = \, \vec{0} 
\; , \quad 
\nabla \cdot \vec{B} \, = \, 0 
\; , \quad 
\nabla \times \vec{\mathcal{H}} \, = \, \vec{0} 
\; , \quad 
\nabla \cdot \vec{D} \, = \, q \, \delta ^{(3)} \big( \vec{r} \big) 
\; ,
\end{equation}
where $q$ is the charge and $\delta ^{(3)}$ is the three-dimensional
Dirac delta distribution. While the curl equations are satisfied by any
spherically symmetric $\vec{E}$ and $\vec{\mathcal{H}}$ fields, the 
divergence equations determine the spherically symmetric $\vec{D}$ 
and $\vec{B}$ fields uniquely,
\begin{equation}\label{eq:DBstat}
\vec{D} 
= \dfrac{q}{4 \, \pi \, r^2} \, \vec{e}{}_r
\; , \qquad 
\vec{B} = 0
\; ,
\end{equation}
where $\vec{e}{}_r$ is the radial unit vector. So whatever the
constitutive law may be, the electric excitation $\vec{D}$ always
has its standard Coulomb form, i.e., it diverges like $r^{-2}$ if
the position of the charge is approached, and the magnetic field
strength $\vec{B}$ vanishes everywhere. The corresponding 
(spherically symmetric) electric field strength $\vec{E}$ and
magnetic excitation $\vec{\mathcal{H}}$ are 
not restricted by Maxwell's equations; they have to be determined 
from the constitutive law.

\subsection{Standard Maxwell vacuum theory}
\label{subsec:Maxstat}
In the standard Maxwell vacuum theory the constitutive
law simply requires $\vec{E}=\vec{D}$ and $\vec{B}= \vec{\mathcal{H}}$. 
Hence, (\ref{eq:DBstat}) says that $\vec{E}$ is the standard Coulomb 
field and that $\vec{\mathcal{H}}$ vanishes, 
\begin{equation}\label{eq:EMaxstat}
\vec{E} = \dfrac{q}{4 \pi r^2} \, \vec{e}{}_r
\, , \qquad
\vec{\mathcal{H}} = \vec{0} \, .
\end{equation}
Clearly, $\big| \vec{E} \big|$ becomes infinite at the origin, 
i.e., at the position of the charge. The direction of $\vec{E}$
is always radial, so in the limit $r \to 0$ the direction may
be any unit vector depending on how the origin is approached.
As these direction vectors average to zero, the Lorentz
force ($\sim \vec{E}$) exerted by the static particle onto itself 
vanishes. Here we follow the widely accepted hypothesis that the
self-force results from averaging over directions, cf. e.g. 
Poisson, Pound and Vega~\cite{PoissonPoundVega:2011}. This 
hypothesis is very natural if one thinks of the point particle
as being the limiting case of an extended (spherical) body.   
 
The field energy in a ball $K_R$ of radius $R$ around the origin is
\begin{equation}\label{eq:Maxstatenergy}
W (R) \, = \, \int _{K_R}  
\dfrac{1}{2} \, \vec{D} 
\cdot \vec{E} \, r^2 \, \mathrm{sin} \, \vartheta \, dr \, d \vartheta
\, d \varphi  = \dfrac{q^2}{8 \, \pi} \,  \int _0 ^R   
 \dfrac{dr}{r^2} \, .
\end{equation}
Clearly, this expression is infinite, for arbitrarily small $R$. 
Both $\vec{E}$ and $\vec{D}$ throw in a factor of $r^{-2}$; one of 
them is killed by a factor of $r^2$ from the volume element but the 
other one makes the integral diverge. We see that we can cure this 
infinity by introducing a modified constitutive law that leaves 
$\vec{E}$ bounded if the position of the charge is approached. We 
will now verify that both the Born-Infeld theory and the Bopp-Podolsky 
theory have this desired property. 

\subsection{Born-Infeld theory}
\label{subsec:BIstat}
\begin{figure}[h]
\begin{center}
    \psfrag{x}{$r$} 
    \psfrag{y}{$\,$ \hspace{-0.6cm} $\big| \vec{E} \big|$} 
    \psfrag{a}{$\,$ \hspace{-0.4cm} $b \:$} 
\includegraphics[height=5.1cm]{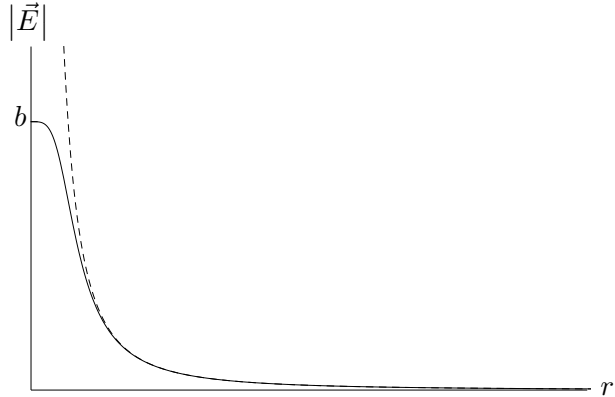}
\end{center}
\caption{Modulus of the electric field strength for a static 
       charge in the Born-Infeld theory (solid) and in the 
       standard Maxwell vacuum theory (dashed)}
\label{fig:BIstat}
\end{figure}

As $\vec{B} = \vec{0}$ by (\ref{eq:DBstat}), in the Born-Infeld theory 
the constitutive law requires 
\begin{equation}\label{eq:DEBIstat}
\vec{D} \, = \, 
\dfrac{\vec{E}}{\sqrt{\, 1 \, - \, \dfrac{1}{b^2} \, 
\big| \vec{E} \big|^2 \,}}
\, , \qquad 
\vec{\mathcal{H}} = \vec{0} \, . 
\end{equation}
With $\vec{D}$ given by (\ref{eq:DBstat}), we have to solve the
equation
\begin{equation}\label{eq:DEBIstat2}
\dfrac{\vec{E}}{\sqrt{\, 1 \, - \, \dfrac{1}{b^2} \, 
\big| \vec{E} \big|^2 \,}}
\, = \, \dfrac{q}{4 \, \pi \, r^2} \, \vec{e}{}_r
\end{equation}
for $\vec{E}$ to determine the electric field strength.
The result is (Born and Infeld~\cite{BornInfeld:1934})
\begin{equation}\label{eq:EBIstat}
\vec{E} =  \dfrac{q}{4 \, \pi \, \sqrt{r_0^4+r^4}} 
\, \vec{e}{}_r \; , \qquad
r_0^2 = \dfrac{q}{4 \, \pi \, b} \, .
\end{equation}
Hence $| \vec{E} |  \to b$ for $r \to 0$, see Fig.~\ref{fig:BIstat}. 
Note that the limit of $\vec{E}$ for $r \to 0$ does not exist because 
the \emph{direction} of the limit vector depends on how the 
position of the point charge is approached. One may say that 
the electric field strength stays bounded but has a 
\emph{directional singularity} at the origin. By averaging
over directions, the self-force ($\sim \vec{E}$) of the static 
particle vanishes.

The field energy in a ball 
$K_R$ of radius $R$ around the origin is
\begin{equation}\label{eq:BIstatenergy}
W (R) \, = \, \int _{K_R}  
 \dfrac{1}{2} \, \vec{D} 
\cdot \vec{E} \, r^2 \, \mathrm{sin} \, \vartheta \, dr \, d \vartheta
\, d \varphi  = \dfrac{q^2}{8 \, \pi \, } \int _0 ^R   
 \dfrac{dr}{\sqrt{r_0 ^4+r^4}} \, .
\end{equation}
This is an elliptic integral which is finite as long as $r_0 >0$, i.e., 
as long as $b$ is finite. Even the field energy in the whole space is finite,
\begin{equation}\label{eq:BIstatenergy2}
\underset{R \to \infty}{\mathrm{lim}} W (R) \, = \, 
\dfrac{q^2 \, \Gamma (5/4)^2}{2 \, r_0 \, \pi ^{3/2}} \, ,
\end{equation}
where $\Gamma$ is the Euler gamma function.

\subsection{Bopp-Podolsky theory}
\label{subsec:BPstat}
\begin{figure}[h]
\begin{center}
    \psfrag{x}{$r$} 
    \psfrag{y}{$\,$ \hspace{-0.6cm} $\big| \vec{E} \big|$} 
    \psfrag{a}{$\,$ \hspace{-1.1cm} $\dfrac{q}{8 \pi \ell ^2} \:$} 
\includegraphics[height=5.5cm]{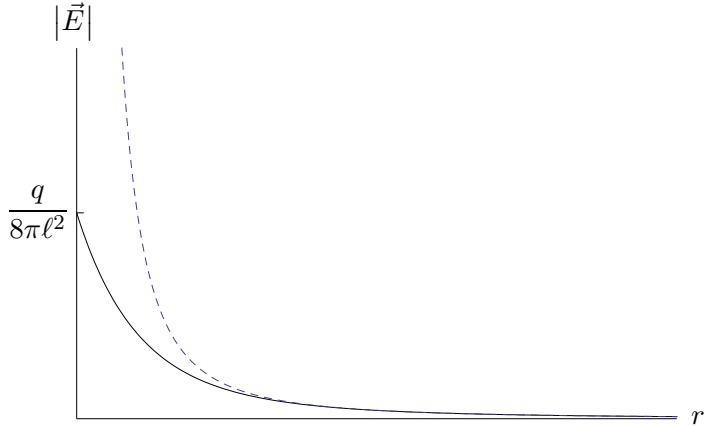}
\end{center}
\caption{Modulus of the electric field strength for a static 
       charge in the Bopp-Podolsky theory (solid) and in the 
       standard Maxwell vacuum theory (dashed)}
\label{fig:BPstat}
\end{figure}

In this case the constitutive law requires
\begin{equation}\label{eq:DEBPstat}
\vec{D} \, = \, \vec{E} 
\, - \, \ell ^2 \, \Delta  \vec{E} 
\, , \quad
\vec{\mathcal{H}} = \vec{0}
\, .
\end{equation}
With $\vec{D}$ given by (\ref{eq:DBstat}), we have to 
solve the second-order differential equation 
\begin{equation}\label{eq:DEBPstat2}
\vec{E} 
\, - \, \ell ^2 \, \Delta  \vec{E} 
\, = \, \dfrac{q}{4 \, \pi \, r^2} \, \vec{e}{}_r
\end{equation}
to determine $\vec{E}$. For a spherically symmetric
field, $\vec{E} \big( \vec{r} \big) = E(r) \vec{e}{}_r 
\big( \vec{r} \big)$, this reduces to
\begin{equation}\label{eq:DEBPstat3}
E \, - \, \dfrac{\ell ^2}{r^2} \, 
\left( \dfrac{d}{dr} \Big( r^2 \dfrac{dE}{dr} \Big)
\, - \, 2 \, E \right) 
\, = \, \dfrac{q}{4 \, \pi \, r^2} \, .
\end{equation}
The general solution is 
\begin{equation}\label{eq:EBPstat}
\vec{E} = \dfrac{q}{4 \, \pi \, r^2} \Big\{ \, 1 \,  
+ C_1 \, \ell \, (r- \ell) \, e^{ r / \ell} 
-  C_2 \, \ell \, (r+ \ell ) \, e^{- r / \ell}
\Big\} \, \vec{e}{}_r 
\end{equation}
with two integration constants $C_1$ and $C_2$. The first 
integration constant is fixed if we require $\vec{E}$ 
to fall off towards infinity; this yields $C_1=0$. The
second integration constant is fixed if we require $\vec{E}$ 
to stay bounded if the position of the charge is approached;
this yields $C_2= \ell {}^{-2}$. This gives us the 
Bopp-Podolsky analogue of the Coulomb $\vec{E}$ field 
(Bopp~\cite{Bopp:1940}, Podolsky~\cite{Podolsky:1942})
\begin{equation}\label{eq:EBPstat2}
\vec{E} = \dfrac{q}{4 \, \pi \, r^2} \Big\{ \, 1 \,  
-  \Big( \dfrac{r}{\ell} + 1 \Big) \, e^{- r / \ell}
\Big\} \, \vec{e}{}_r 
\end{equation}
which satisfies $\big| \vec{E} \big| \to q/(8 \pi \ell {}^2)$
for $r \to 0$, see Fig.~\ref{fig:BPstat}. Just as in the Born-Infeld 
case, the electric field strength stays bounded but has a directional 
singularity at the origin.  

The field energy in a ball 
$K_R$ of radius R around the origin is
\begin{gather}\label{eq:BPstatenergy}
W(R) \, = \, \int _{K_R}  
\dfrac{1}{2} \, \vec{D} 
\cdot \vec{E} \, r^2 \, \mathrm{sin} \, \vartheta \, dr \, d \vartheta
\, d \varphi  
\\
\nonumber
= \,
\dfrac{q^2}{8 \, \pi} \int _0 ^R   
\left(\dfrac{1}{r^2} - \dfrac{e^{-r / \ell}}{r^2} - 
\dfrac{e^{- r / \ell}}{r \ell}
\, 
\right) dr
\, = \, 
\dfrac{q^2}{8 \pi} \, \left(
\dfrac{1}{\ell} \, + \, \dfrac{e^{-R/\ell}-1}{R} \right)
\end{gather}
which is finite as long as $\ell >0$. As in the Born-Infeld theory,
even the field energy in the whole space is finite,
\begin{equation}\label{eq:BPstatenergy2}
\underset{R \to \infty}{\mathrm{lim}} W(R) \, = \, 
\dfrac{q^2}{8 \, \pi \, \ell} \, .
\end{equation}

\section{Field of an accelerated point charge}
\label{subsec:acc}
We have seen that both the Born-Infeld theory and the 
Bopp-Podolsky theory modify the Coulomb $\vec{E}$ field
of a point charge \emph{at rest} in such a way that $| \vec{E}|$
is bounded and that the field energy in a ball around the charge 
is finite. Of course, what one is really interested in is the
field produced by an \emph{accelerated} charge. We will
now try to find out what can be said about this case.

We choose an inertial coordinate system on 
Minkowski spacetime, $g = \eta _{ab} \, dx^a \otimes dx^b$. We
fix a timelike $C^{\infty}$ curve $z^a (\tau )$ 
parametrised by proper time, 
\begin{equation}\label{eq:proper}
\eta _{ab} \dot{z}{}^a 
\dot{z}{}^b = -1 \, .
\end{equation}
We assume that this timelike curve is inextendible. As an 
accelerated worldline may reach (past or future) infinity 
in a finite proper time, this does not necessarily mean 
that the parameter $\tau$ ranges over all of $\mathbb{R}$.
We denote the interval on which $\tau$ is defined by 
$\, ] \, \tau _{\mathrm{min}}, \tau _{\mathrm{max}} \, [ \,$
where $- \infty \le \tau _{\mathrm{min}} < \tau _{\mathrm{max}}
\le \infty$.
   
\begin{figure}[h]
\begin{center}
    \psfrag{x}{$x$} 
    \psfrag{z}{$\quad$ $z (\tau )$} 
    \psfrag{v}{$e_0(\tau)$} 
    \psfrag{e}{$e_{\mu}(\tau)$} 
    \psfrag{t0}{$\theta ^0$} 
    \psfrag{t1}{$\theta ^{\mu}$} 
\includegraphics[height=7.5cm]{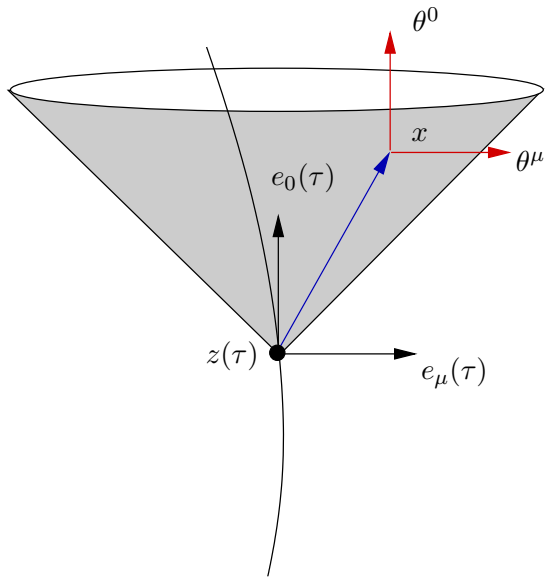}
\end{center}
\caption{Retarded light-cone coordinates and orthonormal coframe}
\label{fig:tetrad}
\end{figure}

We want to determine the electromagnetic field of a point
charge that moves on the worldline $z^a ( \tau )$.
For convenience, we introduce an orthonormal tetrad 
$\big( e_0 (\tau ), e_1 (\tau ), e_2 (\tau ), e_3 (\tau ) \big)$
along the worldline of the charged particle such that
\begin{equation}\label{eq:tetradprop}
e_0^a (\tau ) = 
\dot{z}{}^a (\tau ) 
\, , \qquad
a (\tau) e_3^b (\tau ) = 
\ddot{z}{}^b (\tau ) 
\, ,
\end{equation}
see Fig.~\ref{fig:tetrad}. Along the worldline, this fixes the timelike 
vector $e_0 (\tau )$ everywhere and the spacelike vector $e_3 ( \tau )$, 
up to sign, at all events where the acceleration $\ddot{z} ( \tau )$ is 
non-zero. $e_1 ( \tau )$ and $e_2 ( \tau )$ are then fixed up to a rotation in 
the plane perpendicular to $e_3 ( \tau )$. At points where the acceleration 
is zero, $e_3 ( \tau )$ is ambiguous; there are pathological cases where it
is not possible to extend it into such points such that the resulting
vector field $e_3 $ is continuously differentiable. We exclude such cases
in the following and assume that the tetrad is smoothly dependent on $\tau$ 
and satisfies (\ref{eq:tetradprop}) along the entire worldline.
   
With respect to this tetrad, we can introduce 
retarded light-cone coordinates $(\tau , r , \vartheta , \varphi )$
which are related to the inertial coordinates $(x^0,x^1,
x^2,x^3)$ by
\begin{equation}\label{eq:nuc}
x^a =z^a ( \tau )+r \Big( \dot{z}{}^a ( \tau ) + 
n^a ( \tau , \vartheta, \varphi ) \Big)
\end{equation}
where 
\begin{equation}\label{eq:defn}
n^a ( \tau, \vartheta, \varphi)  = 
\mathrm{cos} \, \varphi \;
\mathrm{sin} \, \vartheta \, e_1^a ( \tau )
\, + \,
\mathrm{sin} \, \varphi \;
\mathrm{sin} \, \vartheta \, e_2^a ( \tau )
+
\mathrm{cos} \, \vartheta \, e_3^a ( \tau )
\, .
\end{equation}

Retarded light-cone coordinates are routinely used nowadays when
treating self-force problems, cf. e.g. Poisson, Pound and
Vega~\cite{PoissonPoundVega:2011}. These coordinates have 
a long history. In connection with electrodynamics on Minkowski 
spacetime, they were introduced by Newman and Unti~\cite{NewmanUnti:1963}  
in 1963. In particular, Newman and Unti demonstrated that in these 
coordinates the Li{\'e}nard-Wiechert potential takes a surprisingly 
simple form. In general relativity the history of light-cone 
coordinates is even older. They made their first appearance in a 1938 
paper by Temple~\cite{Temple:1938} who called the time-reversed
version (i.e., the advanced light-cone coordinates) ``optical 
coordinates''. Advanced light-cone coordinates are used in gravitational 
lensing and in cosmology where the wordline is interpreted as an 
observer who receives light (rather than as a source that emits 
radiation).
 
\begin{figure}[h]
\begin{center}
    \psfrag{x}{$x$} 
    \psfrag{z}{$z (\tau )$} 
\includegraphics[height=5.5cm]{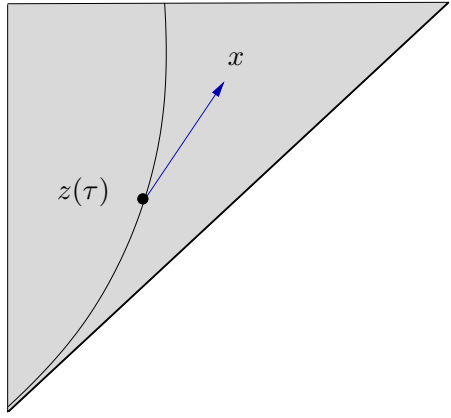}
\end{center}
\caption{Domain of definition, $U$, of the retarded 
light-cone coordinates}
\label{fig:future}
\end{figure}

In retarded light-cone coordinates, the ``temporal'' coordinate
$\tau$ labels the future light-cones with vertex on the chosen
worldline; $r$ is a radius coordinate along each light-cone and
$(\vartheta , \varphi )$ are standard spherical coordinates that
parametrise the spheres $(\tau , r) = \mathrm{constant}$. Of
course, there are the usual coordinate singularities of the
spherical coordinates at the poles $\mathrm{sin} \, \vartheta =0$
and $\varphi$ is defined only modulo $2 \pi$. If these coordinate
singularities are understood, the system of retarded light-cone 
coordinates is well-defined on an open subset, $U$, which
equals the causal future of the worldline with the worldline 
itself being omitted. Fig.~\ref{fig:future} shows a worldline 
that approaches the speed of light in the past. In this case
the causal future of the worldline is bounded by a lightlike 
hyperplane to which the worldline is asymptotic. For a worldline
that does not approach the speed of light in the past, the causal
future is all of Minkowski spacetime. (Recall that we consider
only wordlines that are inextendible.)

With the retarded light-cone coordinates 
$(\tau , r , \vartheta , \varphi )$
we can associate an orthonormal coframe 
$( \theta ^0 ,  \theta ^1 , \theta ^2 , \theta ^3 )$ 
defined by 

\begin{gather}\label{eq:coframe}
\theta ^0   = d \tau + dr + 
r \, a(\tau ) \mathrm{cos} \, \vartheta \, d \tau
\\[0.01cm]
\nonumber
\theta ^1   =  dr + r \, a( \tau ) \mathrm{cos} \, \vartheta
 \, d \tau
\\[0.01cm]
\nonumber
\theta ^2   = r \, d \vartheta -  r  \, a( \tau) \, \mathrm{sin}
\, \vartheta \, d \tau
\\[0.01cm]
\nonumber
\theta ^3   = r \, \mathrm{sin} \, \vartheta \, d \varphi \, ,
\end{gather}
see Fig.~\ref{fig:tetrad}.

The electromagnetic field of the point charge is to be determined by 
solving the Maxwell equations (\ref{eq:Max}) with
\begin{equation}\label{eq:jpoint}
^* \! j (x)
\, = \, 
q \left( \int _{\tau _{\mathrm{min}}} ^{\tau _{\mathrm{max}}}
\delta ^{(4)} \big( x - z ( \tau) \big) \, 
\dot{z}^a ( \tau ) \, d \tau \right) \, \eta _{ab} \, dx^b  
\end{equation}
where $\delta ^{(4)}$ is the 4-dimensional Dirac delta distribution.
The solution has to satisfy the vacuum constitutive law on the
open domain $U$ and it should be retarded. By the latter 
requirement we mean that the field strength at an event $x \in U$ 
should be completely determined by what the point charge did 
in the causal past of the event $x$. 

\subsection{Standard Maxwell vacuum theory}
\label{subsec:MaxLW}
In the case of the standard Maxwell vacuum theory, finding
the field of a point charge on Minkowski spacetime is a 
standard text-book matter. The solution is 
$F=dA, \: H = {}^*{\!}F$, where 
\begin{equation}\label{eq:LW}
A \, = \, - \, \dfrac{q \, \theta ^0}{4 \, \pi \, r}
\, = \, - \, \dfrac{ q}{4 \, \pi} \, 
\Big( 
\, \dfrac{d \tau + dr}{r} 
\, + \, a( \tau) \mathrm{cos} \, \vartheta \,  d \tau 
\Big) 
\end{equation}
is the (retarded) Li{\'e}nard-Wiechert potential. At an 
event $x \in U$, the potential is determined by the 4-velocity 
and the 4-acceleration of the point charge at the retarded time 
which is given by the coodinate $\tau$. There are no ``tail terms'', 
i.e., there is no dependence on the earlier history of the 
point charge.   

For deriving the Li{\' e}nard-Wiechert potential in a systematic
way, one introduces the potential, $F=dA$, and imposes the Lorenz
gauge condition, $d ^* \! A = 0$. Then the first Maxwell equation,
$dF=0$, is automatically satisfied and the second Maxwell equation,
$dH = j$, becomes an inhomogeneous wave equation for $A$,
\begin{equation}\label{eq:waveA}
\square A = \big( {}^*{\!} d ^* {\!} d +
d ^* {\!} d ^* {\!} \big) A = 
 {}^*{\!} d ^* {\!} F=
{}^* {\!} dH = {}^* {\!} j
\, .
\end{equation}
With the well-known (retarded) Green function of the wave operator 
$\square$, the (retarded) solution can be written as an integral
over $^* {} \!j$. Inserting the current from (\ref{eq:jpoint}) gives the 
desired result. 

From the Li{\' e}nard-Wiechert potential we find that the field 
strength $F=dA$ and the excitation $H={}^*{\!}dA$ are given by
\begin{gather}\label{eq:LWF}
F = 
\dfrac{q}{4 \pi} \left( \dfrac{ \theta ^1 \wedge \theta ^0}{r^2}
+ \dfrac{a ( \tau )}{r} \mathrm{sin} \, \vartheta \, \theta ^2 \wedge
(\theta ^0 - \theta ^1 ) \right)
\nonumber
\\
=
\dfrac{q}{4 \pi} \left( \dfrac{dr \wedge d \tau}{r^2}
+ a( \tau ) \,  \mathrm{sin} \, \vartheta \, d \vartheta \wedge d \tau 
\right)
\end{gather}
and
\begin{gather}\label{eq:LWH}
H = 
\dfrac{q}{4 \pi} \left( \dfrac{ \theta ^2 \wedge \theta ^3}{r^2}
- \dfrac{a ( \tau )}{r} \mathrm{sin} \, \vartheta \, \theta ^3 \wedge 
(\theta ^0 - \theta ^1 ) \right)
\nonumber
\\
=
\dfrac{q}{4 \pi} \,
\mathrm{sin} \, \vartheta \, d \vartheta \wedge d \varphi
\, ,
\end{gather}
respectively. Decomposing into electric and magnetic parts 
yields
\begin{gather}\label{eq:LWEBDH}
E_{\mu} \theta ^{\mu} = D_{\mu} \theta ^{\mu} = \dfrac{q}{4 \pi} \, 
\Big\{ \dfrac{\theta ^1}{r^2} + a( \tau ) \, \mathrm{sin} \, \vartheta \, 
\dfrac{\theta ^2}{r} \Big\}
\; , 
\\
B_{\mu} \theta ^{\mu} = {\mathcal{H}} _{\mu} \theta ^{\mu} = 
\dfrac{q}{4 \pi} \, a( \tau ) \, \mathrm{sin} \, \vartheta \, 
\dfrac{\theta ^3}{r} \, .
\end{gather}
In addition to the ``Coulomb part'', which goes with $1/r^2$, we have
in the case of a non-vanishing acceleration a ``radiation part''
which goes with $1/r$. The self-force, i.e. the Lorentz force
exterted onto the point charge by its own field, is given as the 
limit of $q E_{\mu} \theta ^{\mu}$ if the position of the point charge
is approached. The Coulomb part averages to zero, as in the case of a 
static charge. The radiation part, however, does not average to zero;
it gives an infinite self-force whenever the acceleration $a( \tau )$
is non-zero. As in the static case, the field energy in an 
arbitrarily small sphere around the point charge is infinite. 
It is this infinite amount of energy carried by the point charge 
with itself that makes mass renormalisation necessary if one 
wants to formulate an equation of motion for the point charge
taking the self-force into account.

\subsection{Born-Infeld theory}
\label{subsec:BILW}
If one wants to find the field of an accelerated point
charge in the Born-Infeld theory, one would try to mimic
the derivation of the Li{\'e}nard-Wiechert potential as
far as possible. As in the standard Maxwell theory, one
can satisfy the first Maxwell equation by introducing
the potential and one can impose the Lorenz gauge 
condition (or any other gauge condition if this appears
to be more appropriate). However, with $H$ given in 
terms of $F=dA$ by the Born-Infeld constitutive law, 
the second Maxwell equation now becomes a \emph{non-linear} 
inhomogeneous wave equation for $A$. There are no standard
methods for solving such an equation; in particular, 
Green function methods are not applicable. Therefore, 
we cannot write down a Born-Infeld analogue of the
Li{\'e}nard-Wiechert potential. In the Born-Infeld 
theory, no explicit solution of the electromagnetic field 
of a point charge with non-vanishing acceleration seems 
to be known.

One might say that it is not actually necessary to 
write down a solution explicitly. It would be sufficient
if one could verify some properties of the solution. 
Firstly, it would be highly
desirable to prove that, for a point charge moving on
an arbitrary worldline or on a worldline subject 
to some conditions, the retarded electromagnetic field
is unique and regular on $U$. Secondly, it would be
highly desirable to know if for this solution the
self-force and the energy in a ball around the charge
are finite. However, very little is known about these
issues in the Born-Infeld theory beyond the case of an
unaccelerated point charge.

As to regularity, it sems worthwile to point out that
even for a \emph{time-independent} and smooth $j$
the question of regularity is highly non-trivial. It was
shown only recently by Kiessling~\cite{Kiessling:2011}
that in this case the electromagnetic field is, indeed,
free of singularities or discontinuities. Although this
result seems to be intuitively quite obvious, the proof
is difficult and very technical. It is based on series
expansions with respect to $1/b^2$, where $b$ is the
Born-Infeld constant, and the hard part is in the 
proof of convergence. For the field of an accelerated
point charge, it is very well conceivable that infinities
or discontinuities (``shocks'') are formed. It is true
that Boillat~\cite{Boillat:1970} has shown the 
non-existence of some kind of shocks in the Born-Infeld
theory, but these results do not apply to the case at
hand where the equations become singular along a worldline.
 
Even if it is possible to show that the electromagnetic
field of a point charge is regular on $U$, either for
all worldlines or for a special class of worldlines,
it is far from obvious that the field has the same behaviour
as in the static case if the position of the point charge
is approached. A discussion of related issues can be 
found in a paper by Chru{\'s}ci{\'n}ski~\cite{Chruscinski:1998};
this, however, is based on the \emph{assumption} that the
electric field strength remains bounded and that the 
electric excitation diverges like $r^{-2}$ if the position of 
the point charge is approached. In contrast to the 
retarded light-cone coordinates used here, Chru{\'s}ci{\'n}ski
used Fermi coordinates in a similar fashion as they had
been used already earlier by Kijowski~\cite{Kijowski:1994}
in the context of the standard Maxwell vacuum theory.

Something can be said, at least, for the case of a point
charge that is initially at rest and then starts 
accelerating. In this case, conservation of energy 
guarantees that the total field energy must be finite
for all times. However, even in this case it is not clear 
if shocks are excluded.
 
For approaching the problem in a systematic way, one 
may write the electromagnetic field strength as a 
power series with respect to $1/b^2$,
\begin{equation}\label{eq:BIFN}
F \, = \, \sum _{N=0} ^{\infty} \, \dfrac{F_N}{b^{2N}}  \,= \,
F_0 \, + \, \dfrac{F_1}{b^2} \, + \, \dots ,
\qquad F_N = dA_N \, .
\end{equation}
Inserting this expression into the Born-Infeld 
constitutive law (\ref{eq:BIcon}) and collecting
terms of equal powers of $1/b^2$ gives 
\begin{equation}\label{eq:BIHN}
H \, = \, \sum _{N=0} ^{\infty} \, \dfrac{H_N}{b^{2N}}  \,= \,
\sum _{N=0} ^{\infty} \, \dfrac{1}{b^{2N}} \,
\left( \, {}^*{\!}F_N \, + \,
 \mathcal{W}_N \big( F_0, \dots, F_{N-1} \big) \, \right)   
\end{equation}
where $\mathcal{W}_N \big( F_0, \dots, F_{N-1} \big)$ stands
for an expression depending on $F_0 , \dots, F_{N-1}$ that
can be explicitly calculated for every $N$.  
We have to determine the $F_N = dA_N$ such that $dH =j$ with
the current given by (\ref{eq:jpoint}). This can be done
by requiring
\begin{equation}\label{eq:BIHhier}
dH_0 = j \, , \quad dH_N = 0 \: \, \text{for} \: \, N=1,2, \dots 
\end{equation}
and solving these equations iteratively. We may impose the 
Lorenz gauge condition on each $A_N$. Then the zeroth order
retarded solution is known to be the standard Li{\'e}nard-Wiechert
field, $F_0=dA_0$ with $A_0$ given by the right-hand side of
(\ref{eq:LW}). The higher-order $F_N=dA_N$ are determined by
\begin{equation}\label{eq:BIAhier}
d \left( {}^*{\!}dA_N \, + \, 
 \mathcal{W}_N \big( dA_0, \dots, dA_{N-1} \big) \, \right) = 0
\, .
\end{equation}
In the Lorenz gauge, this is the standard inhomogeneous
wave equation for $A_N$, with the inhomogeneity given in
terms of the lower-order solutions $A_0, \dots , A_{N-1}$,
\begin{equation}\label{eq:boxA}
\square A_N = {}^*{\!} \tilde{j}{}_{N} \, , \qquad
\tilde{j}{}_N = - d \mathcal{W}_N \big( dA_0, \dots, dA_{N-1} \big) \, .
\end{equation}
The retarded solution of this equation is known from classical 
electrodynamics: It is the retarded potential of the ``current'' 
three-form $\tilde{j}{}_N$. In this way, we can iteratively determine
the $A_N$ and write the solution $F=dA$ as a formal power
series.

The big question, unanswered so far, is whether or not
this series converges. We do know that it does converge 
in the case of vanishing acceleration; then we get the
field of a static point charge discussed in Section 
\ref{subsec:BIstat}. For non-zero acceleration, however, 
no convergence results are known.

\subsection{Bopp-Podolsky theory}
\label{subsec:BPLW}
In the case of the Bopp-Podolsky theory the situation 
is much better than in the case of the Born-Infeld
theory. The Bopp-Podolsky theory is linear, so it allows
for applications of the Green function method.

With $F=dA$ and choosing the Lorenz gauge, $d ^*{\!}A = 0$,
the remaining field equation reads
\begin{equation}\label{eq:BPwA}
\square A-  \ell ^2 \, \square ^2 A \, = \,{}^*{\!} j \; .
\end{equation}
This fourth-order equation for $A$ can be reduced to a pair of 
second-order equations 
\begin{equation}\label{eq:hAtA}
\square \hat{A} = {}^*{\!} j \, , \qquad
\square \tilde{A} - \ell ^{-2} \tilde{A} = j \, ,
\end{equation}
if we write
\begin{equation}\label{eq:hAtA1}
A=\hat{A}-\tilde{A}
\end{equation}
\begin{equation}\label{eq:hAtA2}
\hat{A}:=A-\ell ^2 \square A \, , \qquad
\tilde{A}:=-\ell ^2 \square A \, .
\end{equation}
If rewritten in this way, a quantised version of the theory
would predict the existence of a massless photon, described by 
$\hat{A}$, and a massive photon with Compton wave-length $\ell$,
described by $\tilde{A}$. Both Bopp~\cite{Bopp:1940} and 
Podolsky~\cite{Podolsky:1942} had realised that their
higher-order theory can be rewritten in this way as  a 
two-field theory. This two-field theory is precisely what
Land{\'e} and Thomas~\cite{LandeThomas:1941} independently 
suggested one year after Bopp and one year before Podolsky.

\begin{figure}[h]
\begin{center}
    \psfrag{x}{$x$} 
    \psfrag{z}{$z (\tau )$} 
    \psfrag{q}{$z (\tau ')$} 
    \psfrag{v}{$e_0(\tau)$} 
    \psfrag{e}{$e_{\mu}(\tau)$} 
\includegraphics[height=7.5cm]{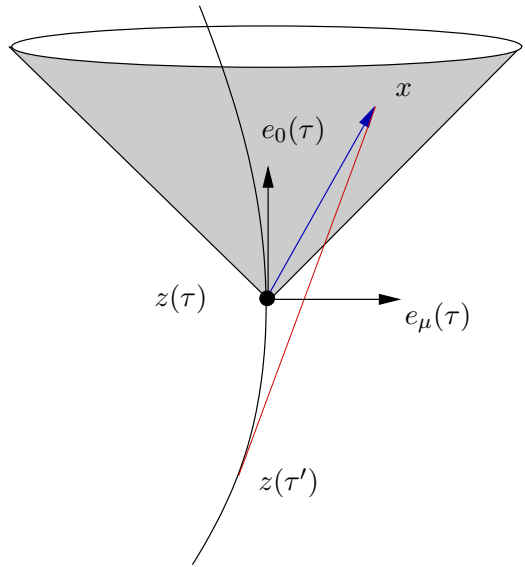}
\end{center}
\caption{$s(x, \tau ')$ is the Lorentz length of the timelike 
line segment that connects $x$ with $z ( \tau ' )$}
\label{fig:s}
\end{figure}

One can thus construct the (retarded) solution to the 
fourth-order equation (\ref{eq:BPwA}) from the (retarded)
Green functions of the wave equations (\ref{eq:hAtA}). 
The latter are well known, see e.g. the original paper 
by Land{\'e} and Thomas~\cite{LandeThomas:1941}.
This gives the retarded solution to (\ref{eq:BPwA})
for the current (\ref{eq:jpoint}) of a point charge as
\begin{equation}\label{eq:BPA}
A (x) \, = \, \left( \int _{- \infty} ^{\tau} 
\dfrac{J_1 \big( s( x, \tau ') /  \ell \big)}{ \ell \, s( x, \tau ' ))}
\dot{z}{}^a ( \tau ' ) \, d \tau ' \right) \, \eta _{ab} \, dx^b 
\end{equation}
where 
\begin{equation}\label{eq:s}
s ( x, \tau ')^2 \, = \, 
- \, \big( x^a - z ^a ( \tau ') \big) \,
\big( x_a - z _a ( \tau ') \big)
\end{equation}
and $J_1$ is the Bessel function of the first kind.
The geometric meaning of 
$s(x, \tau ')$ is illustrated in Fig.~\ref{fig:s}.

The potential (\ref{eq:BPA}) is the Bopp-Podolsky analogue 
of the Li{\'e}nard-Wiechert potential. In contrast to the 
standard Li{\'e}nard-Wiechert potential, it depends on the 
entire earlier history of the point charge up to the retarded 
time $\tau$. Such ``tail terms'' are nothing peculiar; they 
also occur in the standard vacuum Maxwell theory on a 
curved background, see e.g. Poisson, Pound and 
Vega~\cite{PoissonPoundVega:2011}. The integral in 
(\ref{eq:BPA}) and in the corresponding expression for
the field strength can be expanded in a formal power series
with respect to $\ell$. For the self-force, after averaging
over directions this results in a series with terms of 
order $\ell ^{-1}$, $\ell ^0$, $\ell$, $\ell^2 \dots$,
see Zayats~\cite{Zayats:2013} (also cf. McManus~\cite{McManus:1948},
Frenkel~\cite{Frenkel:1996} and Frenkel and 
Santos~\cite{FrenkelSantos:1999}). However, these series are
non-convergent and, therefore, of limited use. 

So in contrast to the Born-Infeld theory, in the 
Bopp-Podolsky theory the electromagnetic potential
(and, thereupon, the electromagnetic field strength) 
produced by an arbitrarily accelerated point charge
can be explicitly written down, albeit in terms of 
an integral over the particle's earlier history.
A detailed discussion of the class of worldlines
for which this integral absolutely converges will 
be given elsewhere~\cite{GratusPerlickTucker:2014}.
This demonstrates that, for a large class of worldlines, 
the electric field stays bounded and there is no
need for mass renormalisation. 
As an important example, the self-force of a 
\emph{uniformly} accelerated point charge was
calculated by Zayats~\cite{Zayats:2013}. 

Because of the tail terms, the equation of motion is
no longer a differential equation but rather an 
integro-differential equation for the worldline. It
is unknown if the equation of motion admits run-away
solutions. For some partial results, indicating that
run-away solutions cannot exist if $\ell$ is bigger 
than a certain critical value, see Frenkel and 
Santos~\cite{FrenkelSantos:1999}.

\section{Implications for gravity}
\label{subsec:grav}
The preceding discussion can be summarised in the following way.
In the standard Maxwell vacuum theory, the self-force is infinite
and mass renormalisation is necessary. Postulating a negative
infinite bare mass is conceptually not satisfactory and the
resulting equation of motion, the Lorentz-Dirac equation, is
highly pathological. In the Born-Infeld theory, the properties
of the field of a static charge look promising, but for an
accelerated charge very little can be calculated and the
properties of the field are largely unknown.
For the Bopp-Podolsky theory, the field of an accelerated
point charge can be calculated, in terms of an integral over
the history of the particle which is manageable to a certain
extent, and it can be shown for a large class of accelerated
worldlines that the field is, actually, finite. No negative
infinite bare mass needs to be postulated, and the equation
of motion can be assumed to be the usual Lorentz-force
equation with the (finite) self-field included after averaging
over directions. The explicit
expression of the electromagnetic field, given by the analogue
of the Li{\'e}nard-Wiechert potential, is more complicated
than in the standard vacuum Maxwell theory on Minkowski spacetime,
because of the tail terms. However, such tail terms are familiar
from the standard vacuum Maxwell theory on a curved spacetime and
should not be viewed as a reason for discarding the theory. 
Although there are still several open issues -- most notably the 
absence or non-absence of run-away solutions has to be clarified 
-- it seems fair to say that in the Bopp-Podolsky theory
the infinities associated with point charges are cured to 
a large extent. We may therefore view it as the best candidate 
for a conceptually satisfactory theory of classical charged 
point particles. (This does not necessarily mean that the 
Bopp-Podolsky theory is ``the correct theory of electromagnetism'' 
at a fundamental, quantum field theoretical, level).  

Do these observations teach a lesson with respect to the
gravitational self-force? In the approximation where the
self-gravitating particle is viewed as a perturbation of
a fixed background spacetime, the theory is very similar to 
the electromagnetic case in the standard Maxwell vacuum 
theory. Modifying the theory along the lines of the 
Born-Infeld theory seems to be of no use: Firstly, it
is largely unclear if the Born-Infeld theory really cures 
the infinities in the field of an accelerated point 
charge. Secondly, the original Einstein theory was
already a non-linear theory whose non-linearities had
been killed by setting up the approximation formalism for
the self-gravitating point mass. Therefore, it seems rather
meaningless to re-introduce non-linear terms. The situation
is quite different for the Bopp-Podolsky theory. Here 
linearity is kept but higher-order terms are added. It
seems not unreasonable to assume that Einstein's theory
can be modified by adding higher-order terms in such a
way that they survive the approximation, giving rise 
to a regularising term of the same kind as it occurs
in the Bopp-Podolsky theory. Higher-order theories
of gravity have been investigated intensively. They are
mainly motivated by the observation that quantum corrections
to Einstein's theory are expected to give a Lagrangian
that is of quadratic or higher order in the curvature,
resulting in field equations that involve
fourth derivatives of the metric. (The simplest class of 
such theories is the class of $f(R)$ theories which are 
reviewed, e.g., in the Living Review by de Felice and 
Tsujikawa~\cite{FeliceTsujikawa:2010}). Looking for a version that 
gives rise to a Bopp-Podolsky-like term seems to be a 
promising programme that might give a new theoretical 
framework for getting a finite gravitational self-force. 

\section*{Acknowledgements}
This work was financially supported by the Deutsche 
Forschungsgemeinschaft, Grant LA905/10-1, and by the 
German-Israeli-Foundation, Grant 1078/2009. Moreover,
I gratefully acknowledge support from the Deutsche 
Forschungsgemeinschaft within the Research Training
Group 1620 ``Models of Gravity''. As to the part
on Bopp-Podolsky theory, I wish to thank Robin Tucker
and Jonathan Gratus for many helpful discussions and for 
the ongoing collaboration on this subject.  Finally,
I am grateful to the organisers of the Heraeus-Seminar
``Equations of motion in relativistic gravity'' for 
inviting this contribution.

\bibliographystyle{unsrt}

\bibliography{perlick_eom_proceedings_2013}

\end{document}